\documentclass[aps,prd,reprint,nofootinbib,showkeys,showpacs,superscriptaddress,longbibliography,floatfix]{revtex4-1}

\pdfoutput=1

\usepackage{amsmath}
\usepackage{amssymb}
\usepackage[american]{babel}
\usepackage{booktabs}
\usepackage{dcolumn}  
\usepackage[T1]{fontenc}  
\usepackage{graphicx}  
\usepackage[colorlinks]{hyperref}  
\usepackage[utf8]{inputenc}  
\usepackage{multirow}
\usepackage{txfonts}  
\usepackage{url}
\usepackage[dvipsnames]{xcolor}
\usepackage{float}
\usepackage{lipsum}

\usepackage{tikz}

\usetikzlibrary{arrows,shapes,positioning,shadows,trees}

\tikzset{
	basic/.style  = {draw, text width=2cm, drop shadow, font=\sffamily, rectangle},
	root/.style   = {basic, rounded corners=2pt, thin, align=center,
		fill=green!30},
	level 2/.style = {basic, rounded corners=6pt, thin,align=center, fill=green!60,
		text width=8em},
	level 3/.style = {basic, thin, align=left, fill=pink!60, text width=6.5em}
}

\usetikzlibrary{arrows,decorations.pathmorphing,backgrounds,fit,positioning,shapes.symbols,chains}

\newcolumntype{d}[1]{D{.}{.}{#1}}
\newcolumntype{v}[1]{D{,}{,\ }{#1}}

\makeatletter

\newcommand{\Rmnum}[1]{\expandafter\@slowromancap\romannumeral #1@}
\makeatother

\renewcommand {\arraystretch}{1.3}

\hypersetup{
	citecolor = blue,
	linkcolor = RoyalPurple,
	urlcolor = blue
}

\begin{document}

\title{Thermodynamic Constraints on the Dark Sector}

\author{W. J. C. da Silva}
\email{williamjouse@fisica.ufrn.br}
\affiliation{Departamento de F\'{\i}sica, Universidade Federal do Rio Grande do Norte, Natal, Rio Grande do Norte, 59072-970, Brasil}

\author{J. E. Gonzalez}
\email{gonzalezsjavier@gmail.com}\affiliation{Departamento de F\'{\i}sica, Universidade Federal do Rio Grande do Norte, Natal, Rio Grande do Norte, 59072-970, Brasil} 

\author{R. Silva}
\email{raimundosilva@fisica.ufrn.br}\affiliation{Departamento de F\'{\i}sica, Universidade Federal do Rio Grande do Norte, Natal, Rio Grande do Norte, 59072-970, Brasil}\affiliation{Departamento de F\'{\i}sica, Universidade do Estado do Rio Grande do Norte, Mossor\'o, Rio Grande do Norte, 59610-210, Brasil}

\author{J. S. Alcaniz}
\email{alcaniz@on.br}
\affiliation{Observat\'orio Nacional,Rio de Janeiro, Rio de Janeiro, 20921-400, Brasil}

\pacs{}

\date{\today}

\begin{abstract}
In this paper, we present a unified scheme based on the fluid description of the dark sector of the universe. The scheme captures models with interaction between dark energy and dark matter, being the core of generalization the time-varying equation-of-state parameter $\omega(a)$ and the time-dependent interactions through the interaction function $\epsilon(a)$, where $a$ is the scale factor. Furthermore, we propose thermodynamics constraints on this generalized class of models using the laws of thermodynamics which are combined with observational data. In order to test the observational viability of the unified model, we perform a Bayesian analysis using cosmic chronometers, type Ia supernovae, cosmic microwave background, and angular baryon acoustic oscillation measurements.	
	
\end{abstract}

\maketitle

\section{Introduction}\label{}

Observational and theoretical effort in modern cosmology has successfully revealed a picture of the universe which is very well realized through the standard model of cosmology,  the $\Lambda$ + Cold Dark Matter ($\Lambda$CDM) model. The $\Lambda$ component is responsible for a repulsive gravitational force at cosmological scales which accelerates the universe, whereas the dark matter provides explanation at galaxy and galaxy cluster scales for some gravitational phenomena, e.g., the observed rotation curves of galaxies and structure formation \cite{Aghanim:2018eyx,Alam2017,troxel}. Despite the success of the standard model, our understanding of the universe still lacks a plausible explanation of the theoretical and observational issues associated with the late-time accelerated cosmic expansion of the universe \cite{Buchert2017}.

These open questions have motivated some approaches which can be divided into two classes: extensions of general relativity (GR) \cite{Clifton2012} and dark energy models \cite{Amendola2018}. The former one considers infrared modifications to GR, leading to a weakening of gravity on cosmological scales, whereas the latter adds an exotic component of dark energy, e.g., a scalar field to the r.h.s of Einstein field equations. Within this context, the dark energy component can also be described through a fluid approach, with thermodynamics playing an important role in such a description (see, e.g., \cite{lima-alcaniz2004,Brevik2004,Bousso2005,Izquierdo2006,SILVA2007,Pereira2008,Pavn2008} and references therein). 

Thermodynamic considerations have been combined with observational data to provide physical constraints on the fluid dark energy (see, e.g.,~\cite{heydson2012,Sharov2016,Nunes2016,Gimenes2018,gonzalez2018,Contreras2018,daSilva2019,Silva2019}). In this paper, we concentrate on a framework which generalizes scenarios based on the following assumptions: (i) a perfect fluid of dark energy with constant equation of state (EoS) \cite{lima-alcaniz2004}, (ii) an imperfect fluid (bulk viscosity) of dark energy with a varying EoS and a non-null chemical potential \cite{heydson2013}, (iii) interaction among the dark components with a constant parameter of interaction, being a viscous fluid representing the dark energy as well as a pressureless fluid, as a typical cold dark matter fluid \cite{gonzalez2018}, (iv) a coupling model with both constants EoS and interaction parameter \cite{ernandes2009}, and (v) with the time-dependent interaction rate \cite{ernandes2010}. 

From a phenomenological perspective, we propose a unified scheme based on the fluid description of the dark sector which is also able to recover the models discussed above. Mathematically speaking, the core of the interacting models proposed here is based on a time varying equation-of-state (EoS) parameter  $\omega(a)$ and the time-dependent interactions through the interaction function $\epsilon(a)$. Moreover, we obtain thermodynamics constraints on this generalized class of models using the laws of thermodynamics which are combined with observational data. The observational viability of the model considered will be discussed by performing a Bayesian  analysis using cosmic chronometers (CC), type Ia supernovae (SNe Ia), cosmic microwave background (CMB), and angular baryon acoustic oscillation (BAO) measurements.

The paper is organized as follows. In Sect. \ref{sec:general-model}, we present our generalized interacting model. In Sect. \ref{sec:thermo-constraints}, based on the second law of thermodynamics, we derive thermodynamic constraints on the model parameters. In Sect. \ref{sec:data-method}, we show the data considered in this work. Our results are presented in Sect. \ref{sec:results}. In Sect. \ref{sec:conclusions}, our main conclusions are presented.

\section{Generalized Interacting Model}\label{sec:general-model}

Let us consider a homogeneous, isotropic, and flat cosmological background described by Friedmann-Lemaître-Robertson-Walker metric (FRLW) and assume that the cosmic budget is composed for baryons (b), dark matter (dm), radiation (r), and dark energy (de). We treat the dark matter and dark energy as interacting fluids with the energy-momentum tensor of the dark sector given by
\begin{equation}\label{coupled-tensor}
	T_{\mu\nu} = T_{\mu\nu}^{\text{dm}} + T_{\mu\nu}^{\text{de}}.
\end{equation}

The covariant conservation of energy-momentum tensor, $\nabla_{\mu}T^{\mu\nu} = 0$, leads to
\begin{equation}\label{conservation}
\dot{\rho}_\text{dm} + 3H\rho_\text{dm} = -\dot{\rho}_\text{de} - 3H\rho_\text{de}(1+\omega) = Q,
\end{equation}
where $\rho_{\text{dm}}$ and $\rho_{\text{de}}$ represent the energy density of cold dark matter and dark energy, respectively, while $Q$ is the phenomenological interaction term. Note that $Q > 0$ indicates the dark energy decaying into dark matter while $Q < 0$ implies the opposite. 

The evolution of the dark components can be found by solving the system of Eq. (\ref{conservation}). Generally, this can be done by assuming a form for $Q$ \cite{wang-meng2005, alcaniz2005,vonMarttens:2018iav} or by assuming a relation between the energy densities of the components \cite{antonella2018,Carneiro:2019rly}. Since in the standard description, the dark matter density evolves as $\rho_{\text{dm}} \propto a^{-3}$, here, we consider a deviation from the standard evolution characterized by the following function \cite{ernandes2009, ernandes2010}

\begin{equation}\label{dm-evolution}
\rho_{\text{dm}}=\rho_{\text{dm},0}a^{-3+\epsilon \left(a\right)},
\end{equation}
where $\rho_{\text{dm},0}$ is the today dark matter energy density calculated in $a_0 = 1$ and $\epsilon \left(a\right)$ is a function that depends of scale factor. 

In what follows, we consider that the equation of state of dark energy is described by a function of the scale factor, $p = \omega(a)\rho_{\text{de}}$. By considering the equation of state of dark energy, as shown in Eq. (\ref{coupled-tensor}), the dark matter evolution,  as shown in Eq. (\ref{dm-evolution}), and the relation between scale factor and redshift $1/a = 1+z$, we obtain
\begin{equation}\label{density-complete}
	\begin{aligned}
	\rho _{x}  = & \rho _{\text{x},0}\exp{\left[\int 3\frac{\left[ 1+\omega \left( z\right)\right] }{1+z}dz\right]} + \\ & + \frac{\rho _{\text{dm},0}\int \left( 1+z\right) ^{3-\epsilon \left(z\right) }\ln \left( 1+z\right) \epsilon \left( z\right) ^{\prime }\exp{\left[\int 3\frac{\left[ 1+\omega \left( z\right)\right] }{1+z}dz\right]}dz}{\exp{\left[-\int 3\frac{\left[ 1+\omega \left( z\right)\right] }{1+z}dz\right]}} + \\ & + \frac{\rho _{\text{dm},0} \int \epsilon \left( z\right) \left( 1+z\right)^{2-\epsilon \left( z\right)}\exp{\left[\int 3\frac{\left[ 1+\omega \left( z\right)\right] }{1+z}dz\right]}dz}{\exp{\left[-\int 3\frac{\left[ 1+\omega \left( z\right)\right] }{1+z}dz\right]}},
	\end{aligned}%
\end{equation}
where $\rho _{\text{x},0}$ is an integration constant associated with dark energy density,  $\omega(z)$ is the time-dependent EoS parameter of dark energy fluid, and $\epsilon \left(z\right)$ is the interaction function. We will assume that the functional form of the equation-of-state parameter is $\omega(z) = \omega_0 + \omega_zf(z)$. In the literature, many different parameterizations are proposed (see, e.g., \cite{escamilla2016, escamilla2016.2} and references therein). In this work, we use two parameterizations widely discussed in the literature known as Chevallier-Porlaski-Linder parameterization (CPL) \cite{chevallier2000, linder2002} and Barbosa-Alcaniz parameterization (BA) \cite{barboza-alcaniz2008}
\begin{equation}\label{par-wz}
f(z) = \left\{
\begin{array}{lllll}
\frac{z(1+z)}{1+z^{2}}  &(\mbox{BA})
\\
\frac{z}{1+z} &(\mbox{CPL}). 
\end{array}
\right.
\end{equation}
For a complete description, we need the functional form of $\epsilon \left(a\right)$. In Refs. \cite{ernandes2009, ernandes2010}, it was proposed two parameterizations for interaction function. We consider in this work the simplest choice, i.e.
\begin{equation}\label{par-interaction}
	\epsilon = \epsilon_0(1 + z)^{-\delta},
\end{equation}
where $\epsilon_0$ is a positive constant and $\delta$ is a constant which may take both positive and negative values \cite{alcaniz2005}. Then, combining the parameterizations, Eqs. (\ref{par-wz}), and Eq. (\ref{par-interaction}) in Eq. (\ref{density-complete}), we obtain 
\begin{widetext}
\begin{equation}\label{intecrating-fluid}
\rho_{\text{x}} = \rho _{\text{x},0}\left( 1+z\right) ^{3\left( 1+\omega _{0}\right)
}\left( 1+z^{2}\right) ^{\frac{3}{2}\omega _{z}}+\epsilon _{0}\rho _{\text{dm},0}%
\frac{\int \left( 1+z^{2}\right) ^{-\frac{3}{2}\omega _{z}}\left( 1+z\right)
	^{-3\omega _{0}-\epsilon _{0}(1+z)^{-\delta }-\delta -1}\left[ -\delta \ln
	\left( 1+z\right) +1\right] dz}{\left( 1+z\right) ^{-3\left( 1+\omega
		_{0}\right) }\left( 1+z^{2}\right)^{-\frac{3}{2}\omega _{z}}}, \hspace{0.5cm} (\mbox{BA}) 
\end{equation}
\begin{equation}\label{intecrating-fluid2}
\rho_{\text{x}}  = \rho _{\text{x},0}\left( 1+z\right) ^{3\left( 1+\omega _{0}+\omega
	_{z}\right) }e^{-3\omega _{z}\frac{z}{1+z}} + \epsilon _{0}\rho_{\text{dm},0}\frac{%
	\int e^{(3\omega \frac{z}{1+z})}\left( 1+z\right) ^{-\epsilon
		_{0}(1+z)^{-\delta }-3\left( \omega _{0}+\omega _{z}\right) -\delta -1}\left[
	-\delta \ln \left( 1+z\right) +1\right] dz}{\left( 1+z\right) ^{-3\left(
		1+\omega _{0}+\omega _{z}\right) }e^{(3\omega _{z}\frac{z}{1+z})}}, \hspace{0.5cm} (\mbox{CPL})  
\end{equation}
\end{widetext}
Note that, for $\delta=0$, we obtain the model studied in Ref. \cite{gonzalez2018}, whereas for $\delta = 0$ and $\omega_z = 0$ and $\delta = 0$, $w_0 = -1$ and $\omega_z = 0$, we recover the results of \cite{Jesus:2008xi} and \cite{wang-meng2005,alcaniz2005}, respectively. 
Furthermore, for $\epsilon_0 = 0$, the well-known evolution of a  dynamical dark energy \cite{chevallier2000,linder2002,barboza-alcaniz2008} and the $\omega$-fluid description is fully recovered. From Eqs. (\ref{conservation}) and (\ref{dm-evolution}), it is possible to show that parameterization (\ref{par-interaction}) is equivalent to the coupling term
\begin{equation}
 Q = \epsilon_0 H (1 + \ln a ^ \delta)a^\delta \rho_{\rm dm}\;,   
\end{equation}
which reduces to the well-known parameterization $Q = \epsilon_0 H \rho_{\rm dm}$ for $\delta = 0$.

\section{Thermodynamic Constraints}\label{sec:thermo-constraints}
The thermodynamic state of a relativistic simple fluid is characterized for three quantities: the energy momentum tensor $T^{\mu\nu}$, the particle flow vector $N^{\mu}$, and the entropy flux  $S^{\mu}$, defined, respectively, as \cite{weinberg1971,silva2002,heydson2012}
\begin{equation}
N^{\mu} = nu^\mu,
\end{equation}
\begin{equation}
S^{\mu} = n \sigma u^\mu,
\end{equation}
where $n$ is the particle number density and  $\sigma$ the specific entropy. The fundamental equations of motion are obtained from the covariant derivative of energy-momentum tensor (energy's conservation), particle flux (equation of balance for the particle number), and entropy flux (second law of thermodynamics), then,
\begin{equation}
\nabla_{\mu}T^{\mu\nu} = \dot{\rho} + 3H(\rho + \rho) = 0, 
\end{equation}
\begin{equation}
\nabla_{\mu}N^{\mu} = \dot{n} + 3Hn = \Psi, 
\end{equation}
\begin{equation}
\nabla_{\mu}S^{\mu} = \tau \ge 0, 
\end{equation}
where $\rho$, $p$, $\Psi$, and $\tau$ are energy density, pressure, particle source, and entropy source, respectively. By assuming that the interaction between dark matter and dark energy affects only the particle mass (particle number is conserved, $\Psi = 0$), the fluids are composed by variable mass particles \cite{farrar2003}.  The second law of thermodynamics requires that the entropy source be nonnegative. For $\tau = 0$, we have a non-dissipative states (perfect fluid) and $\tau \ge 0$ denotes a dissipative states (imperfect fluid). The relation between temperature, specific entropy, energy density, pressure, and particle number is given by Gibbs equation 
\begin{equation}
	nTd\sigma = d\rho - \frac{\rho + p}{n}dn.
\end{equation}
Following the standard description (\cite{weinberg1971,lima-germano1992,silva2002}), it is possible to show that the temperature law is given by
\begin{equation}
	\frac{\dot{T}}{T} = \left( \frac{\partial p_{0}}{\partial \rho}\right)_n\frac{\dot{n}}{n} + \left( \frac{\partial \Pi}{\partial \rho}\right)_n\frac{\dot{n}}{n}
\end{equation}
where $p_{0}$ is the equilibrium pressure and $\Pi$ is the bulk viscosity pressure. As is well known, the cold dark matter component is pressureless which implies that there is no temperature evolution law for this component. However, the dark energy has pressure, so that its temperature evolution law is important for the thermodynamics analysis.

Eq. (\ref{conservation}) can be rewritten  as $\dot{\rho}_{\text{x}}+3H\left( 1+\omega_{0}\right) \rho _{\text{x}} =-3H\Pi$, with $\Pi$ given by
\begin{equation}\label{bulk}
\Pi = \omega _{z}f\left( z\right) \rho _{\text{x}} + \frac{\epsilon \left( z\right)}{3} \rho _{\text{dm}}+\frac{\epsilon \left( z\right) ^{\prime }}{3}\rho _{\text{dm}}\left(	1+z\right) \ln \left( 1+z\right),
\end{equation}
which mimics a fluid with bulk viscosity \cite{heydson2012, heydson2013}. The bulk viscosity is a sum of a term related to the variable part of the dark energy EoS, one referring to the interaction term and another one coming from the dependence of the interaction parameter with redshift. Note that in the limit of constant interaction, $\epsilon(z) \rightarrow \epsilon_0$, the results obtained in Ref. \cite{gonzalez2018} are fully recovered. In the uncoupled case, the results of Ref. \cite{heydson2012} are retrieved. By assuming only a scalar dissipative process, i.e., bulk viscosity, the entropy source of the interacting dark fluid is given by \cite{silva2002}
\begin{equation}\label{entropy-source}
	S_{;\mu }^{\mu }=-3H\frac{\Pi }{T_{x}}.
\end{equation}
The dark energy temperature is always positive and increasing with universe's expansion \cite{lima-alcaniz2004,heydson2012,heydson2013}. From Eqs. (\ref{bulk}) and  (\ref{entropy-source}), the second law of thermodynamics implies that
\begin{equation}
	\omega_{z}\leqslant \frac{\epsilon \left( z\right) \rho _{\text{dm}}+\rho
		_{\text{dm}}\left( 1+z\right) \ln \left( 1+z\right) \epsilon \left( z\right)
		^{\prime }}{3f\left( z\right) \rho _{x}}.
\end{equation}
Moreover, from Eqs. (\ref{dm-evolution}), (\ref{par-wz}), (\ref{par-interaction}), (\ref{intecrating-fluid}) and (\ref{intecrating-fluid2}) we obtain the following thermodynamic constraints for BA and CPL parameterizations, respectively
\begin{widetext}
\begin{equation}\label{vinc-1}
\omega _{z} \leqslant -\epsilon _{0}\rho _{dm,0}\frac{\left( 1+z\right) ^{3-\epsilon
		_{0}(1+z)^{-\delta }-\delta }\left[ 1-\delta \ln \left( 1+z\right) \right] }{%
	3\frac{z(1+z)}{1+z^{2}}\left[ \rho _{x,0}\left( 1+z\right) ^{3\left(
		1+\omega _{0}\right) }\left( 1+z^{2}\right) ^{\frac{3}{2}\omega
		_{z}}+\epsilon _{0}\rho _{dm,0}\frac{\int \left( 1+z^{2}\right) ^{-\frac{3}{2%
			}\omega _{z}}\left( 1+z\right) ^{-3\omega _{0}-\epsilon _{0}(1+z)^{-\delta
			}-\delta -1}\left[ -\delta \ln \left( 1+z\right) +1\right] dz}{\left(
		1+z\right) ^{-3\left( 1+\omega _{0}\right) }\left( 1+z^{2}\right) ^{-\frac{3%
			}{2}\omega _{z}}}\right] },
\end{equation}
	
\begin{equation}\label{vinc-2}
\omega _{z} \leqslant -\epsilon _{0}\rho _{dm,0}\frac{\left( 1+z\right)
	^{3-\epsilon _{0}(1+z)^{-\delta }-\delta }\left[ 1-\delta \ln \left(
	1+z\right) \right] }{3\frac{z}{1+z}\left[ \rho _{x,0}\left( 1+z\right)
	^{3\left( 1+\omega _{0}+\omega _{z}\right) }e^{(-3\omega _{z}\frac{z}{1+z}%
		)}+\epsilon _{0}\rho _{dm,0}\frac{\int e^{(3\omega \frac{z}{1+z})}\left(
		1+z\right) ^{-\epsilon _{0}(1+z)^{-\delta }-3\left( \omega _{0}+\omega
			_{z}\right) -\delta -1}\left[ -\delta \ln \left( 1+z\right) +1\right] dz}{%
		\left( 1+z\right) ^{-3\left( 1+\omega _{0}+\omega _{z}\right) }e^{(3\omega
			_{z}\frac{z}{1+z})}}\right] },
\end{equation}
\end{widetext}
which clearly is not defined at $z = 0$. On the other hand, considering null chemical potential, the Euler relation can be written as $S/n = (\rho + p)nT$. Thus, from the positiveness of entropy we obtain
\begin{equation}
\rho_{\text{x}}[1 + \omega(z)] \geq 0\,.
\end{equation}
Now, using Eqs. (\ref{par-wz}), (\ref{intecrating-fluid}) and (\ref{intecrating-fluid2}), we find

\begin{equation}\label{vinc-3}
\rho^{\text{BA}}_{\text{x}}\left[ 1 + \omega_0 + \omega_z\frac{z(1+z)}{1+z^{2}}\right]  \geq 0.
\end{equation}

\begin{equation}\label{vinc-4}
	\rho^{\text{CPL}}_{\text{x}}\left[ 1 + \omega_0 + \omega_z\frac{z}{1+z}\right]  \geq 0,
\end{equation}

From these constraints and considering that dark energy density satisfies the weak energy condition, that is $\rho_{\text{x}} \geq 0$, within the redshift interval of interest, a similar constraint is also obtained for the non-interacting model \cite{heydson2012, heydson2013}
\begin{equation}
	[1 + \omega(z)] \geq 0.
\end{equation}
We completely recovered the results of Ref.\cite{heydson2012} for the uncoupled dark energy case.

\section{Cosmological Data}\label{sec:data-method}

In order to investigate the properties of generalized interaction and impose thermodynamic constraints, we perform a Bayesian statistical analysis using different cosmological probes, which are listed as follows:

\begin{itemize}
\item \textbf{BAO 2D}: Clustering measurements that provide the baryon acoustic oscillations (BAO) data are important to break parameter degeneracies from CMB measurements. This probe, which is almost unaltered by uncertainties in the nonlinear evolution of matter density and other systematics errors, is considered as a statistical standard ruler. Thus, this geometrical probe allows to constrain the background evolution of dark energy models. In this work, we consider the angular BAO measurements ($\theta_{\rm BAO}$) from the galaxy distribution of the DR11 \cite{G-Carvalho} and the quasar distribution of the DR12 \cite{E-Carvalho}. 

\item \textbf{CMB}: The CMB is one of the most important observables in cosmology due to our understanding of linear physics as well as its sensibility to cosmological parameters. In this work, we consider the position of the first peak of the CMB temperature power spectrum, $l_1=220\pm 0.5$ \cite{2016}.  We follow the approach presented in Ref. \cite{gonzalez2018}.

\item \textbf{Cosmic Chronometers (CC)}: Another cosmological probe considered in this work is the cosmic chronometer data obtained through the differential age method. The CC allows to determine the Hubble parameter values at different redshifts taking the relative age of passively evolving galaxies \cite{jimenes2001,Simon:2004,Stern:2009,Moresco:2012,Zhang:2012,Moresco:2015,Moresco:2016}. We use the $31$ available measurements of the Hubble parameter in the redshift range $0.07 < z < 1.96$ listed in Ref. \cite{Farooq:2016zwm}.

\item \textbf{Type Ia Supernovae (SNe Ia)}: The type Ia Supernovae data are among the most important measurements in observational cosmology and constitute a principal evidence of the cosmic acceleration. Type Ia Supernovae are considered as standardizable candles, and they are a powerful probe to constraint cosmological parameters, specially the dark energy EoS. The Pantheon compilation is the most recent SNIa sample which consists of $1048$ measurements of apparent magnitude in the redshift range $0.01 < z < 2.3$ \cite{scolnic2018}.
\end{itemize}
\begin{table}[H]
	\renewcommand{\arraystretch}{1.2}
	\renewcommand{\tabcolsep}{0.2cm}
	\centering
	\caption{The table shows the priors on the free parameters of each parameterization. Note that $\mathcal{N}(\mu, \sigma^2)$ means a Gaussian prior with mean $\mu$ and variance $\sigma^2$, and $\mathcal{U}(a, b)$ means uniform prior.}
	\begin{ruledtabular}
		\begin{tabular}{ccc}
			Parameter       & Prior 						   \\ \colrule
			$H_0$[km/s/Mpc]			    & $\mathcal{N}(74.03,1.42)$ 
			\\
			$\Omega_{c}$     & $\mathcal{U}(0.001, 0.99)$   \\
			$\omega_{0} $  	   & $\mathcal{U}(-2.5, 0)$  \\
			$\omega_{a} $  	   & $\mathcal{U}(-5, 3)$  \\
			$ \epsilon_{0}$   & $\mathcal{U}(0, 0.15)$  \\
			$ \delta$ 	   & $\mathcal{U}(0.0, 10)$  	  
		\end{tabular}
		\label{tab:prior}
	\end{ruledtabular}
	
\end{table}

\section{Results}\label{sec:results}

\subsection{Parameter estimation and thermodynamics constraints}

Using the above mentioned cosmological observations, we adopt the nested sampling \cite{skilling2004} method based on a Monte Carlo technique targeted at the efficient calculation of the evidence, yet which allows posterior probability as a by-product. For this, we implement the public package \textsf{MultiNest} \cite{feroz2007, feroz2013, buchner2014} through the \textsf{PyMultiNest} \cite{pymultinest}. To perform this analysis, we choose uniform priors for all parameters except for the Hubble constant for which we consider a Gaussian prior based on the Cepheids/SNe model-independent $H_0$ value \cite{Riess:2019}. These priors are shown in Table \ref{tab:prior}. We fix the radiation density parameter in $\Omega_{\text{r},0} h^2 = 1.698 \Omega_\gamma$ with $\Omega_\gamma = 2.469 \times 10^{-5}h^2$ and the baryon density at the Planck Collaboration value $\Omega_bh^2=0.02237 $  \cite{Aghanim:2018eyx}.

\begin{figure*}[t]
\centering
\includegraphics[width=0.7\linewidth]{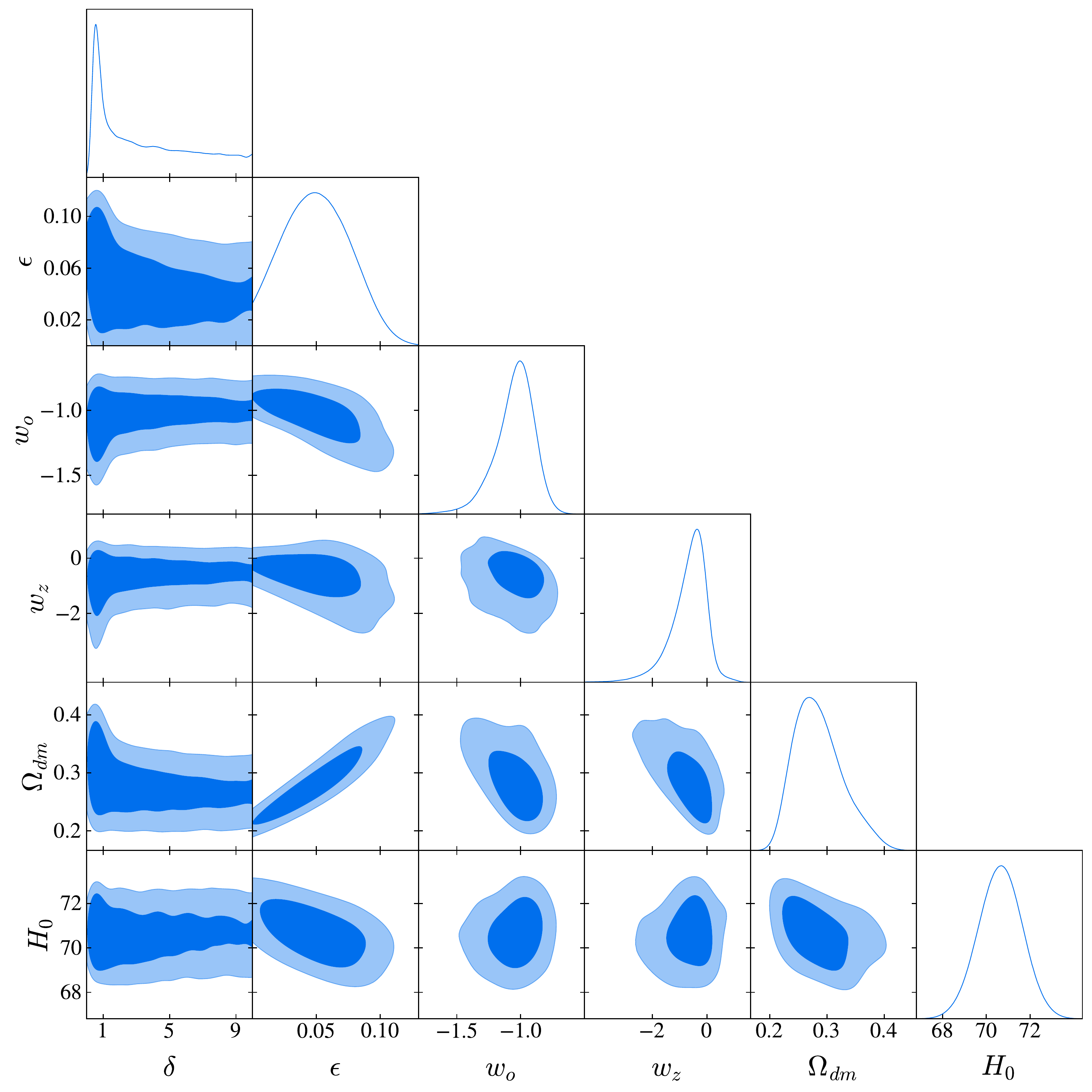}
	\caption{\label{fig:regions1} 1$\sigma$ and 2$\sigma$ confidence regions and the probability density functions for the cosmological parameters constrained by the joint statistical analyses considering the data BAO + CMB + CC + SNe Ia for BA parameterization with a power law interaction.}
\end{figure*}

\begin{figure*}[t]
\centering
	\includegraphics[width=0.7\linewidth]{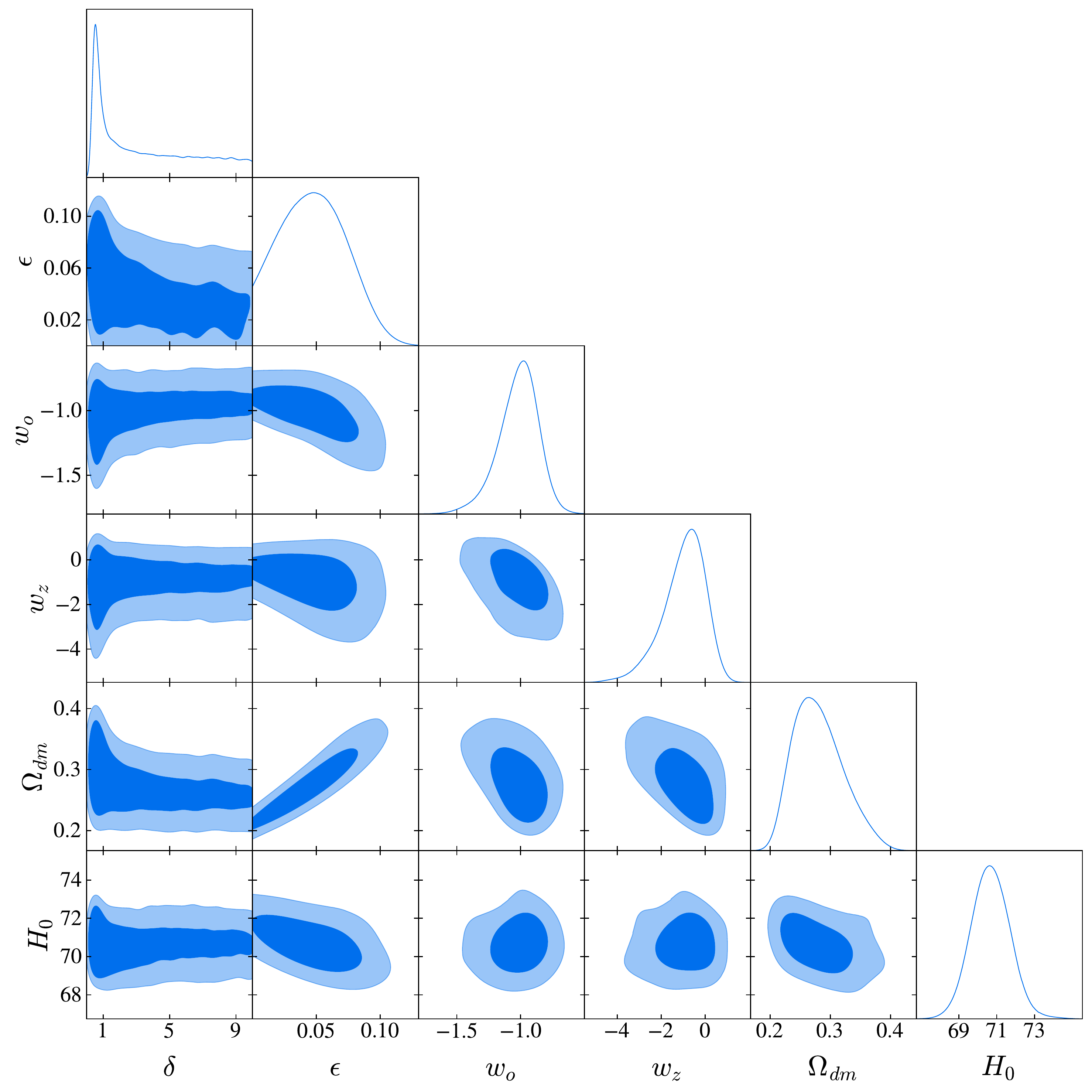}
	\caption{\label{fig:regions3}1$\sigma$ and 2$\sigma$ confidence regions and the probability density functions for the cosmological parameters constrained by the joint statistical analyses considering the data BAO + CMB + CC + SNe Ia for CPL parameterization with a power law interaction.}
\end{figure*}

The results of the joint analysis (BAO + CMB  + CC + SNe Ia) are presented in Table \ref{tab:results}, and in Figs. \ref{fig:regions1}  and \ref{fig:regions3}. Table \ref{tab:results} shows the mean and $1\sigma$ error  for each parameter analyzed. In the Figs. \ref{fig:regions1} and \ref{fig:regions3} we show the posterior distributions and $1\sigma$ and $2\sigma$ contours regions for the parameterizations studied in this work. Note that negative values of $\delta$ are ruled out\footnote{Even if the prior allows negative values of $\delta$, the observational constrains rule them out.} and the current observational bounds on $\delta$ are not restrictive \cite{ernandes2009,ernandes2010}. On the other hand, the values obtained for the interaction parameter $\epsilon_{0}$ are compatible with the results obtained in Refs. \cite{ernandes2009,ernandes2010}. 

We will also combine the thermodynamics bounds discussed earlier with observational data to constrain the $\omega_{0}-\omega_z$ parametric space. We perform a Bayesian analysis with $\epsilon_{0}$ and $\delta$ fixed. By considering the results obtained for BA parameterization, we fix these parameters at the mean value obtained in the global analysis, i.e.,  $\epsilon_{0} = 0.045$ and $\delta = 4.3$.  
For CPL parameterization we consider $\epsilon_{0} = 0.041$ and $\delta = 4.5$. 

\begin{figure*}[t]
    \centering
	\includegraphics[width=8.5cm]{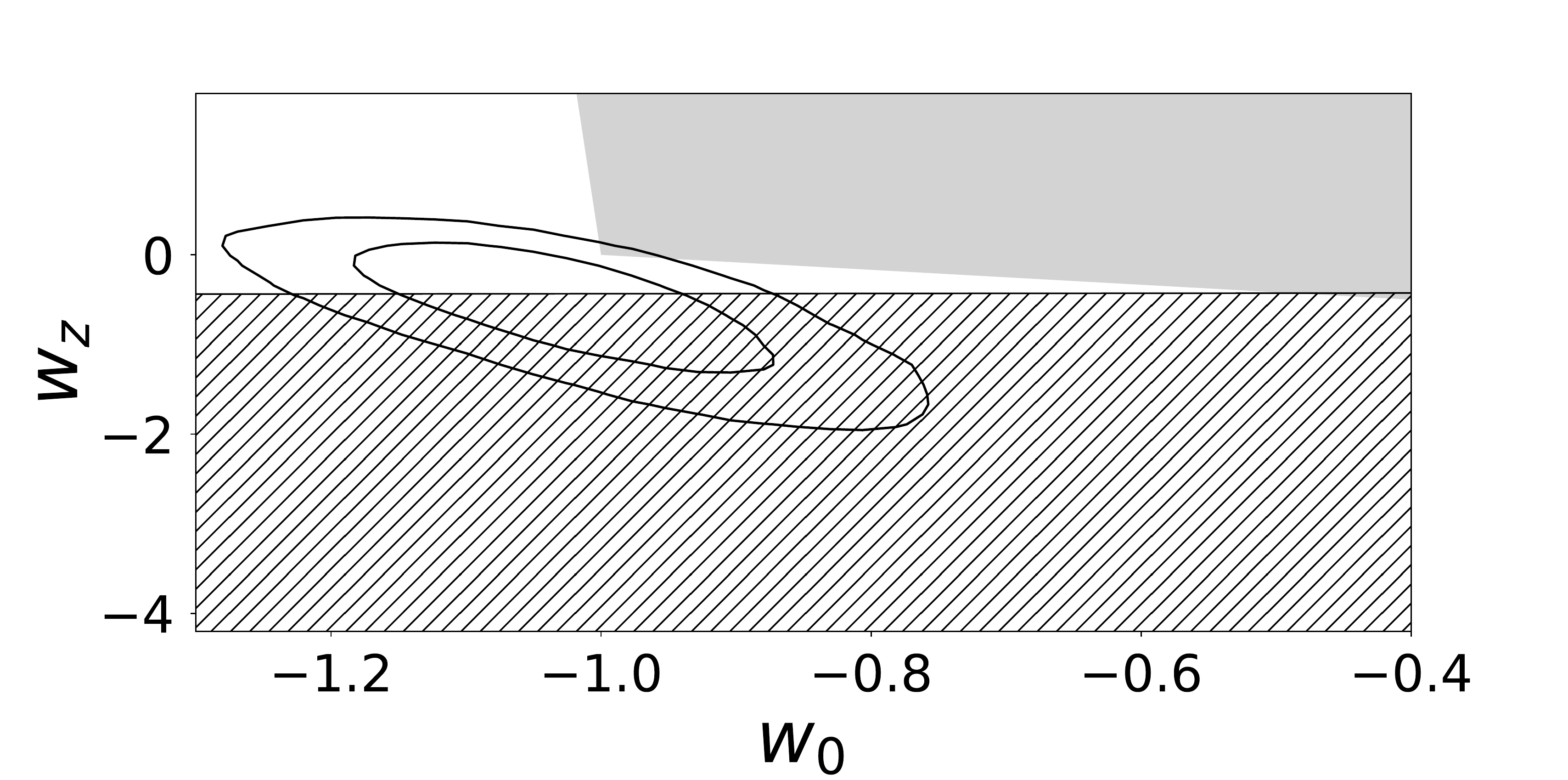}
    \includegraphics[width=8.5cm]{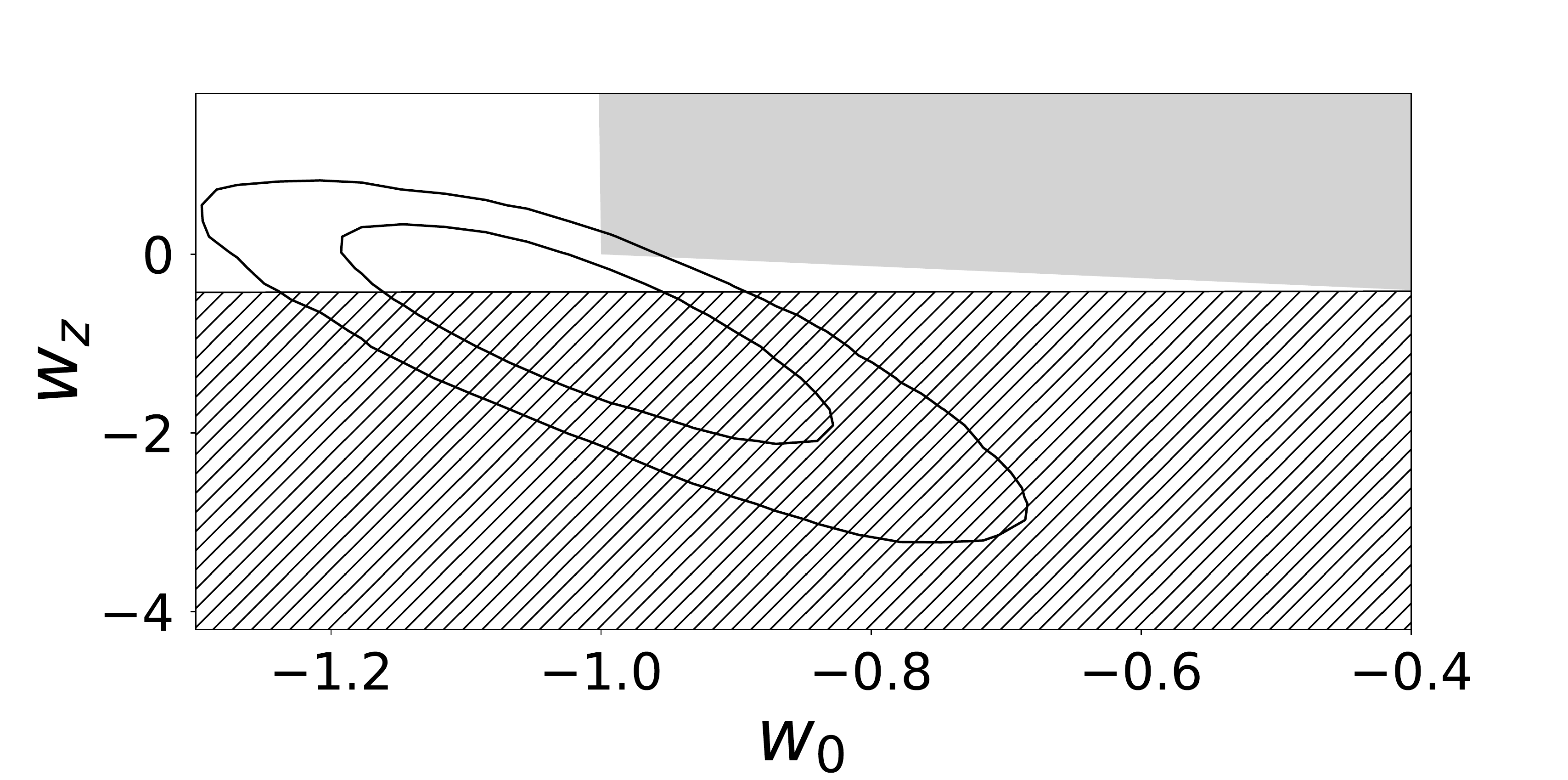}
	\caption{\label{fig:vinc1} Observational and thermodynamics constraints in the $w_o-\omega_z$ plane for the BA (left) and CPL (right) parameterizations. The $\delta$ and $\epsilon$ parameters are fixed at their central value in the global analyses. The hatched region represents the constraints from the second law of thermodynamics, as shown in Eqs. (\ref{vinc-1}) - (\ref{vinc-2}), and gray region means the positiveness of entropy, as shown in Eqs. (\ref{vinc-3}) - (\ref{vinc-4}).} 
\end{figure*}

In Fig. \ref{fig:vinc1} we show the combination between $1\sigma$ and $2\sigma$ confidence contours and the thermodynamics constraints shown in Eqs. (\ref{vinc-1}), (\ref{vinc-2}), (\ref{vinc-3}) and (\ref{vinc-4})  for both parameterizations. 
Since these constraints depend on time, the regions are plotted by assuming its validity within the redshift interval $0.01 < z < 2.3$.

The results are shown in Fig. \ref{fig:vinc1}, where the hatched and gray regions represent the constraints from Eqs. (\ref{vinc-1}) -  (\ref{vinc-2}) and Eqs. (\ref{vinc-3}) - (\ref{vinc-4}), respectively.  Fig. \ref{fig:vinc1}  shows that the observational constraints are more restricted for the BA parameterization; however, in both cases, the thermodynamics and observational constraints are incompatible at the interaction parameters considered, mainly by the higher $\epsilon_0$ value than the one obtained without evolving interaction in Ref. \cite{gonzalez2018}.

\begin{table}
	\centering
	\caption{Statistical constraints on the  cosmological  parameters for each parameterization and the power law model using a Gaussian prior for $H_0$.}
	\begin{ruledtabular}
\begin{tabular} { c c c}
	Parameterization     &  BA  & CPL  \\
   Interaction                      &  Power law&  Power law \\
	\colrule
	$H_0$   &$70.7^{+1.0}_{-1.0}   $  &$70.7^{+1.0}_{-1.0}$\\
	
	$\Omega_{dm}$ & $0.271^{+0.040}_{-0.037}$& $0.263^{+0.039}_{-0.033}$\\
	
	$\omega_0$    & $-1.00^{+0.11}_{-0.13}$ & $-0.98^{+0.12}_{-0.14}$ \\
	
	$\omega_z$    & $-0.51^{+0.44}_{-0.64}$   &    $-0.73^{+0.69}_{-0.99}$ \\
	
	$\epsilon_0$  & $0.045^{ +0.025}_{-0.026}     $  & $0.041^{ +0.026}_{-0.023}     $ \\
	
	$\delta$ & $4.3^{ +4.0}_{-3.5}$  &$4.5^{ +3.9}_{-3.6}$   \\
\end{tabular}
	\label{tab:results}
\end{ruledtabular}
\end{table}

\subsection{Model selection}

In order to compare the several coupling models with $\Lambda$CDM, we implement the Bayesian model comparison in terms of the strength of the evidence according to the Jeffreys scale. To do this, we estimate the values of the logarithm of the Bayesian evidence ($\ln \mathcal{E}$) and the Bayes factor ($\ln \mathcal{B}$). These values were achieved considering the priors defined in the Table \ref{tab:prior} and, the dataset describe in the Sect. \ref{sec:data-method}. We assumed $\Lambda$CDM model as the reference one. The Jeffreys scale interprets the Bayes' factor as follows: inconclusive if $|\ln \mathcal{B}| < 1$, weak if $1 \leq |\ln \mathcal{B}| < 2.5$, moderate if $2.5 \leq |\ln \mathcal{B}| < 5$, and strong if $|\ln \mathcal{B}| \geq 5$. A negative (positive) value for $\ln\mathcal{B}$ indicates that the competing model is disfavored (supported) with respect to the $\Lambda$CDM model.
	
\begin{figure}[H]
	\centering
	\includegraphics[width=8.5cm]{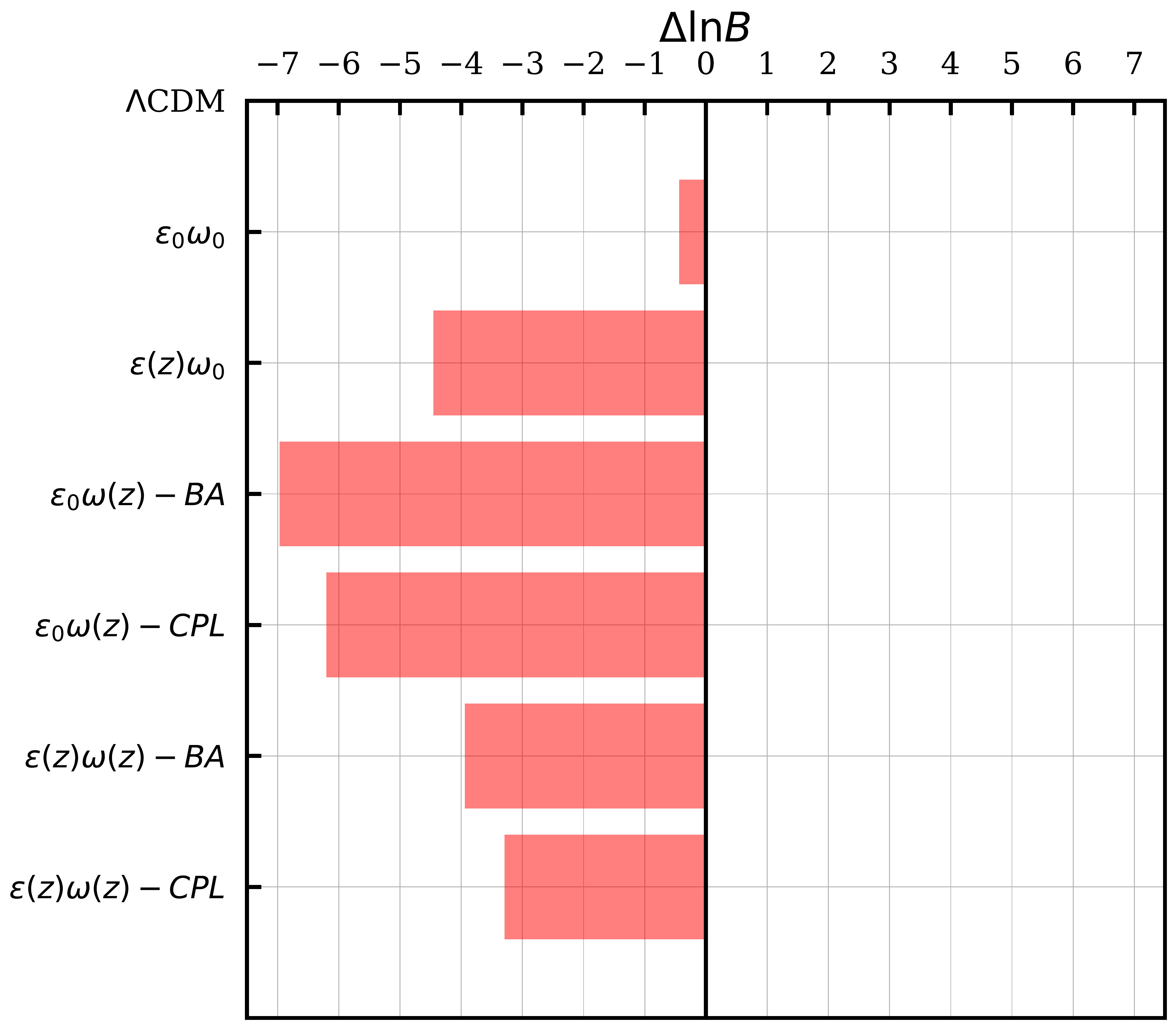}
	\caption {\label{fig:bayes-factor}Figure shows the Bayes factor between the $\Lambda$CDM and each coupling models considering the data combination. The red bar indicates Gaussian prior on $H_{0}$. The coupled models analyzed in this work were: $\epsilon_0\omega_0$ \cite{wang-meng2005}, $\epsilon(z)\omega_0$   \cite{ernandes2009,ernandes2010}, and $\epsilon_{0}\omega(z)$ \cite{gonzalez2018}.    Note that $\Delta\ln \mathcal{B}$ < 0 favors the $\Lambda$CDM.}
\end{figure}

In Fig. \ref{fig:bayes-factor}, we show the values obtained for Bayes factor considering each model studied in this work. Note that all coupling models achieved the negative values for Bayes factor, i.e., the dataset used to perform the statistical analysis prefers the simplest model, $\Lambda$CDM.

\section{Conclusions}\label{sec:conclusions}
In this paper, we proposed the unified scheme following the fluid description of the dark sector of the universe. This generalized interacting model recovered several models proposed in the literature, being the core of the generalization the time-varying equation-of-state (EoS) parameter and the time-dependent interactions, via the interaction function. Based on the
positiveness of the entropy and the second law of thermodynamics, physical constraints were combined with observational ones. Specifically, the bounds on the $\omega(z)$ come from thermodynamics constraints Eqs. (\ref{vinc-1}), (\ref{vinc-2}), (\ref{vinc-3}), and (\ref{vinc-4}) combined with actual observational data BAO + CMB + CC + SNe Ia. We have shown that this combination provided very restrictive limits on the parametric space as shown in Fig. \ref{fig:vinc1}. Finally, in order to investigate the viability of the generalized interacting model, we have performed the Bayesian statistical analysis (see Table \ref{tab:results} and Fig. \ref{fig:bayes-factor}) in order to compare the parameterizations (BA and CPL) used in this generalized approach.

Finally, as mentioned in Sect. \ref{sec:thermo-constraints}, the present work assumes conservation of the number of  particles ($\Psi = 0$). A general approach relaxing this condition is currently under investigation and will be reported in a forthcoming communication.

\begin{acknowledgements}
The authors thank Brazilian scientific and financial support federal agencies, Coordenação de Aperfeicoamento de Pessoal de Nível Superior (CAPES) and Conselho Nacional de  Desenvolvimento  Científico  e  Tecnológico (CNPq). This work was supported by High-Performance Computing Center (NPAD)/UFRN. J. Alcaniz acknowledges support from CNPq (Grants No. 310790/2014-0 and 400471/2014-0) and FAPERJ (Grant No. 233906).
\end{acknowledgements}

\bibliographystyle{apsrev4-1}
\bibliography{references}

\end{document}